\documentclass[12pt]{article}

\usepackage{amsfonts}
\usepackage{amsmath}
\usepackage{amssymb}
\usepackage{graphicx}
\usepackage{color}
\usepackage[normalem]{ulem}

\begin {document}


\title{An LP-based inconsistency monitoring of
pairwise comparison matrices}     

\author{S. Boz\'oki,\ \ J. F\"ul\"op \thanks{Research Group
of Operations Research and Decision Systems,
Computer and Automation Research Institute,
Hungarian Academy of Sciences,
1518 Budapest, P.O. Box 63, Hungary}
\and
W.W. Koczkodaj \thanks{Computer Science, Laurentian University,
Sudbury, Ontario P3E 2C6, Canada}
}
\maketitle

\begin{abstract}
\noindent
A distance-based inconsistency indicator, defined by the third author
for the consistency-driven pairwise comparisons method,
is extended to the incomplete case.
The corresponding optimization problem is transformed
into an equivalent linear programming problem.
The results can be applied in the process
of filling in the matrix as the decision maker
gets automatic feedback. As soon as a serious error
occurs among the matrix elements, even due to a misprint,
a significant increase in the inconsistency index is reported.
The high inconsistency may be alarmed not only at the end
of the process of filling in the matrix
but also during the completion process.
Numerical examples are also provided.
\end{abstract}

\noindent\textbf{Keywords:} linear programming, incomplete data,
inconsistency analysis, pairwise comparisons. \\

\section{Introduction}

The use of pairwise comparisons (PC) is traced by some scholars
to Ramon Llull (1232-–1315). However, it is generally accepted
that the modern use of PC took place in \cite{Saaty1977}.
It is a natural approach
for processing subjectivity although objective data
can be also processed this way.

The consistency-driven approach incorporates
the reasonable assumption
that by finding the most inconsistent assessments,
one is able to reconsider his/her own opinions.
This approach can be extended to incomplete data.

Mathematically, an $n\times n$ real matrix $\mathbf{A}=[a_{ij}]$
is a pairwise comparison (PC) matrix if $a_{ij}>0$ and
$a_{ij}=1/a_{ji}$ for all $i,j=1,\dots ,n$. Elements $a_{ij}$
represent a result of (often subjectively) comparing the $i$th
alternative (or stimuli) with the $j$th alternative according to a
given criterion. A PC matrix $\mathbf{A}$ is consistent if
$a_{ij}a_{jk}=a_{ik}$ for all $i,j,k =1,\dots ,n.$ It is easy to
see that a PC matrix $\mathbf{A}$ is consistent if and only if
there exists a positive $n$-vector $w$ such that $a_{ij}=w_i/w_j,
i,j=1,\dots ,n.$ For a consistent PC matrix $\mathbf{A}$, the
values $w_i$ serve as priorities or implicit weights of the
importance of alternatives.

First, let us look at a simple example of a $3\times 3$ reciprocal matrix:
\begin{equation}
\left(
  \begin{array}{ccc}
    1 & a & b \\
    1/a & 1 & c \\
    1/b & 1/c & 1 \\
  \end{array}
\right). \label{eq:a1}
\end{equation}
Koczkodaj defined the inconsistency index in \cite{Koczkodaj93}
for (\ref{eq:a1}) as
\begin{equation}
CM(a,b,c)= \min \left \{\frac{1}{a}\left| a-\frac{b}{c}\right|,
\frac{1}{b}\left|b-ac\right|, \frac{1}{c}\left| c-\frac{b}{a}\right|
\right\}.\label{eq:a2}
\end{equation}
Duszak and Koczkodaj \cite{DuKo94} extended this definition (\ref{eq:a2})
for a general $n\times n$ reciprocal matrix $\mathbf{A}$ as the
maximum of $CM(a,b,c)$ for all triads $(a,b,c)$,
i.e., $3 \times 3 $ submatrices which are themselves PC matrices,
in $\mathbf{A}$:
\begin{equation}
CM(\mathbf{A})= \max \{CM(a_{ij},a_{ik},a_{jk})\mid 1\le i<j<k\le
n\}. \label{eq:a3}
\end{equation}


The concept of inconsistency index $CM$ is due to the fact that
the consistency of a PC matrix is defined
for (all) triads. By definition, a PC matrix is inconsistent
if and only if it has as least one inconsistent triad. \\

It is shown in \cite{BoRa08} that $CM$ and some other inconsistency
indices are directly related to each other but only in case
of $3 \times 3$ PC matrices. As the size of PC matrix gets larger
than $3 \times 3,$ this function-like relation between different
inconsistency indices does not hold. \\

Incomplete PC matrices were defined by Harker                   
\cite{Harker1987a,Harker1987b}. Harker's main justification of  
introducing incomplete PC matrices is the possible claim for    
reducing the large number of comparisons in case of an e.g.,    
$9 \times 9$ matrix (\cite{Harker1987b}, p.838.). \\            

In the incomplete PC matrix below,
the missing elements are denoted by $\ast$:
\begin{equation} \mathbf{A}=
\begin{pmatrix}
     1     &    a_{12}  &   \ast   & \ldots & a_{1n}   \\
1/{a_{12}} &       1    &  a_{23}  & \ldots & \ast \\
    \ast   & 1/{a_{23}} &     1    & \ldots & a_{3n}   \\
    \vdots &    \vdots  &  \vdots  & \ddots & \vdots   \\
1/{a_{1n}} &    \ast    &1/{a_{3n}}& \ldots &   1    \\
\end{pmatrix}.\nonumber
\end{equation}
We assume that all the main diagonal elements are given and equal to 1.
Incomplete matrices were analyzed
in \cite{BoFuRo2010, FeGi07, KoHeOr99, KwiesielewiczVanUden2003}.
Most previous solutions are based on approaches
for deriving weights from complete PC matrices.
Weighting is not in the focus of the paper,                           
however, incomplete PC matrices are also used in our approach.        


In our model, the use of incomplete PC matrix is rather means than object.  
The aim of the paper is to provide an LP based                              
monitoring system which is able to compute $CM$-inconsistency               
in each step of the filling in process as well as                           
to inform the decision maker if s/he exceeds a given                        
inconsistency threshold. The algorithm is constructive                      
in the sense that it localizes the main root of $CM$-inconsistency.         
It is of fundamental importance to assume that the                          
decision maker should be guided or supported                                
but not led by making the comparisons for him/her in                        
a mechanical way by proposing a reduction algorithm.                        

Decision support tools for controlling or predicting
inconsistency during the filling in process have been provided by
Wedley \cite{Wedley1993},
Ishizaka and Lusti \cite{IshizakaLusti2003}
and Temesi \cite{Temesi2010}.
\\

As the decision maker fills in a PC matrix, an incomplete PC        
matrix is resulted in by adding each element (except for the        
$n(n-1)/2$-th one, then the PC matrix becomes complete),            
whose $CM$-inconsistency, to be defined, plays an            
important role in our inconsistency monitoring system. \\           

Let us assume that the number of the missing elements above the main diagonal
in $\mathbf{A}$ is $d$, hence the total number of
missing elements in $\mathbf{A}$ is $2d$.
Let $(i_l, j_l), l=1,\dots, d$, where $i_l<j_l$,
denote the positions of  the missing elements
above the diagonal in $\mathbf{A}$.

Let us substitute variables $x_{i_lj_l}, l=1,\dots ,d$, for the
missing values above the main diagonal in $\mathbf{A}$. Similarly,
missing reciprocal values below the main diagonal are replaced by
$1/x_{i_lj_l}, l=1,\dots ,d$. We denote the new matrix by
$\mathbf{A}(x_{i_1j_1}, \dots , x_{i_dj_d})$ to attack the following
optimization problem:
\begin{equation}
\begin{array}{lll}
\min &&CM(\mathbf{A}(x_{i_1j_1}, \dots ,
x_{i_dj_d}))\\
{\rm s.t.} &&x_{i_lj_l}>0,\ l=1,\dots ,d,
\end{array}
\label{eq:a4}
\end{equation}
for replacing the missing values by positives values and
their reciprocals so that $CM(\mathbf{A}(x_{i_1j_1}, \dots ,
x_{i_dj_d}))$ is minimal.
It is easy to see the practical ramifications of this approach.
There is no need to continue the filling-in process
of the missing elements
if the optimal value of (\ref{eq:a4}) exceeds
a pre-determined inconsistency threshold.
Instead, one should concentrate on finding
the sources of inconsistency among the
already given elements in $\mathbf{A}$.

\section{The LP form of the optimization problem}

The nonlinear structure of PC matrices is due to the    
reciprocal property ($a_{ij}=1/a_{ji}$).                
However, when the element-wise                          
logarithm of a PC matrix is taken, some properties,     
e.g. the transitivity rule ($a_{ij}a_{jk} = a_{ik}$)   
becomes linear. Linearized forms also occur in          
weighting methods \cite{GoldenWang1990}. \\             

Let $L=\{(i,j)\mid 1 \le i<j \le n, a_{ij} \textrm{\ is\ given}\}$
and $\bar L=\{(i_l,j_l)\mid l=1,\dots,d\}$ denote the index sets of
the given and missing elements, respectively, above the main diagonal
in $\mathbf{A}$. Then, for the sake of a unified
notation, introducing variables $x_{ij}$ for all $(i,j)\in L$ as
well, and an auxiliary variable $u$, problem (\ref{eq:a4}) can be
written into the following equivalent form:
\begin{equation}
\begin{array}{lll}
\min &&u\\
{\rm s.t.} &&CM(x_{ij},x_{ik},x_{jk})\le u,\ 1\le i<j<k\le n,\\
 &&x_{ij}=a_{ij},\ (i,j)\in L,\\
  &&x_{ij}>0,\ (i,j)\in \bar L.
\end{array}
\label{eq:a5}
\end{equation}
It is clear that (\ref{eq:a5}) has a feasible solution, variable $u$
is nonnegative for any feasible solution, the optimal values of
(\ref{eq:a4}) and (\ref{eq:a5}) coincide, furthermore, $x_{ij}>0,
(i,j)\in \bar L$, is an optimal solution of (\ref{eq:a4}) if and
only if it is a part of an optimal solution of (\ref{eq:a5}).

The problem (\ref{eq:a5}) is not easy to solve
directly because of the non-convexity of $CM(\cdot)$ of (\ref{eq:a2})
in the arguments. However, a useful property
of different inconsistency indicators
for $3\times 3$ PC matrices was published in \cite{BoRa08}.
For the $3\times 3$ PC matrix of (\ref{eq:a1}),
 let us denote
\begin{equation}
T(a,b,c)= \max\left \{\frac{ac}{b}, \frac{b}{ac}\right
\}.\label{eq:a6}
\end{equation}
As shown in \cite{BoRa08},
\begin{equation}
CM(a,b,c)=1-\frac{1}{T(a,b,c)},\ \
T(a,b,c)=\frac{1}{1-CM(a,b,c)}.
\label{eq:a7}
\end{equation}

Since the univariate function $f(u)=1/(1-u)$ is strictly increasing
on $(-\infty,1)$ and $T(a,b,c)=f(CM(a,b,c))$, problem (\ref{eq:a5})
can be transcribed into the equivalent form
\begin{equation}
\begin{array}{lll}
\min &&f(u)\\
{\rm s.t.} &&T(x_{ij},x_{ik},x_{jk})\le f(u),\ 1\le i<j<k\le n,\\
 &&x_{ij}=a_{ij},\ (i,j)\in L,\\
  &&x_{ij}>0,\ (i,j)\in \bar L.
\end{array}
\label{eq:a8}
\end{equation}
By substitution $t=f(u)$ and applying the definition (\ref{eq:a6}),
problem (\ref{eq:a8}) can be written into the next equivalent form:
\begin{equation}
\begin{array}{lll}
\min &&t\\
{\rm s.t.} &&x_{ij}x_{jk}/x_{ik}\le t,\ 1\le i<j<k\le n,\\
&&x_{ik}/(x_{ij}x_{jk})\le t,\ 1\le i<j<k\le n,\\
 &&x_{ij}=a_{ij},\ (i,j)\in L,\\
  &&x_{ij}>0,\ (i,j)\in \bar L.
\end{array}
\label{eq:a9}
\end{equation}
Since $t\ge 1$ for any feasible solution of (\ref{eq:a9}), the
variables $x_{ij}$ are positive, and the function $\log t$ is
strictly increasing over $t>0$, we can use the old trick of
the logarithmic mapping:
\begin{equation}
\begin{array}{lll}
&&z = \log t,\\
&&y_{ij}=\log x_{ij},\ 1\le i<j\le n,\\
 &&b_{ij}=\log a_{ij},\ (i,j)\in L.
\end{array}
\label{eq:a10}
\end{equation}
Then (\ref{eq:a9}) can be written into the following equivalent
form:
\begin{equation}
\begin{array}{lll}
\min &&z\\
{\rm s.t.} &&y_{ij}+y_{jk}-y_{ik}\le z,\ 1\le i<j<k\le n,\\
&& -y_{ij}-y_{jk}+y_{ik}\le z,\ 1\le i<j<k\le n,\\
 &&y_{ij}=b_{ij},\ (i,j)\in L.
\end{array}
\label{eq:a11}
\end{equation}
The optimization problem (\ref{eq:a11}) is a linear programming
problem.
It has a feasible solution with the non-negative objective function
over the feasible region.
Consequently, (\ref{eq:a11}) has an optimal solution.
Furthermore, problem (\ref{eq:a4}) has also an optimal solution
because of the chain of equivalent problems established up to this point.
The following statements summarize the
essence of the equivalent transcriptions applied above.

\medskip
\noindent\textbf{Proposition 1.} Problems (\ref{eq:a4}) and
(\ref{eq:a11}) have optimal solutions.

If $\bar x_{ij}>0,(i,j)\in \bar L$, is an optimal solution and $\bar
u$ is the optimal value of (\ref{eq:a4}), then
\begin{equation}
\begin{array}{lll}
&&\bar z = \log \frac{1}{1-\bar u}=-\log (1-\bar u),\\
&&\bar y_{ij}=\log \bar x_{ij},\ \ (i,j)\in \bar L,\\
 &&\bar y_{ij}=\log a_{ij},\ (i,j)\in L,
\end{array}
\label{eq:a12}
\end{equation}
is an optimal solution of (\ref{eq:a11}).

Conversely, if $\bar z$, $\bar y_{ij}, 1\le i<j\le n$, is an
optimal solution of (\ref{eq:a11}), then
\begin{equation}
\begin{array}{lll}
&&\bar x_{ij}= e^{\bar y_{ij}}, \ (i,j)\in \bar L,
\end{array}
\label{eq:a13}
\end{equation}
is an optimal solution and $\bar u=1-e^{-\bar z}$ is the optimal
value of (\ref{eq:a4}). $\hfill\Box$ \\

The most inconsistent triad, which is not necessarily unique,
is identified from the active constraints in  (\ref{eq:a11}).
If $\bar u$ is greater than a pre-defined threshold of
CM inconsistency, then the decision maker  is recommended to
reconsider the triad(s) associated with the active constraints.
This case is involved in the third numerical example in Section 3.


\section{Numerical examples}

Three numerical examples are provided in this section. Let
$\mathbf{A}$ be a $4 \times 4$ incomplete PC matrix as follows:

\[
\mathbf{A} =
\begin{pmatrix}
  1    &   \ast  &   3.5  &       5      \\
  \ast &    1    &    3   &      2.5     \\
 1/3.5 &   1/3   &    1   &       \ast   \\
  1/5  &  1/2.5  &   \ast &       1      \\
\end{pmatrix}
\]

There are no complete triads in $\mathbf{A}$. The optimization
problem (\ref{eq:a4}) can be written for this example as:
\begin{equation}
\begin{array}{lll}
\min &&CM(\mathbf{A}(x_{13}, x_{24}))\\
{\rm s.t.} &&x_{13}, x_{24}>0.
\end{array}
\label{eq:a4example}
\end{equation}


Reformulate (\ref{eq:a4example}) in the same way as (\ref{eq:a11})
is derived from (\ref{eq:a4}), the LP has a unique solution. Apply
(\ref{eq:a12})-(\ref{eq:a13}), $CM^{\ast} = 0.236$
is resulted in as the optimal value of (\ref{eq:a4example}) and
it is still less than the acceptable inconsistency threshold,
assumed by Koczkodaj in \cite{Koczkodaj93} as $CM \leq 1/3$.
Consequently, acceptable inconsistency can still be reached by a
suitable filling-in of the missing elements of $\mathbf{A}.$ \\

\bigskip

Let $\mathbf{B}$ be a $5 \times 5$ incomplete PC matrix as follows:

\[
\mathbf{B} =
\begin{pmatrix}
      1    &    \ast    &    1.5     &      2     &    \ast        \\
    \ast   &      1     &    1/2     &    \ast    &      4         \\
  1/1.5    &      2     &      1     &    \ast    &    \ast        \\
  1/2      &    \ast    &    \ast    &     1      &    1/3         \\
    \ast   &    1/4     &    \ast    &     3      &      1
\end{pmatrix}
\]


Now $CM^{\ast} = 0.62$. The above example with incomplete data
demonstrates there is no way of completing it with data to bring
the inconsistency below $1/3$ hence no need to even collect the
missing data. This may be a helpful way of eliminating data
collection which may be sometimes time consuming hence expensive.
It is also noted that matrix $\mathbf{B}$ does not seem to be so
bad at first sight. Let $C_1, C_2, C_3, C_4, C_5$ denote
the criteria whose importancess are presented in
$\mathbf{B},$ and let $C_i \succ C_j$ denote that '$C_i$ is more important
than $C_j$'. One can check that the pairwise comparisons of
$\mathbf{B}$ reflect ordinally transitive relations:
$C_1 \succ C_3 \succ C_2 \succ C_5 \succ C_4.$
The roots of inconsistency are of cardinal nature rather than ordinal. \\

Let $\mathbf{D}$ be a $7 \times 7$ incomplete PC matrix                      
filled in in a sequential order                                                         
$d_{12}, d_{13}, \ldots, d_{23}, d_{24}, \ldots.$                                       
Assume that the decision maker intends to write $d_{45} = 4$
but s/he happens to mistype it by $d_{45} = \mathbf{1/4}.$
The optimization problem(\ref{eq:a11}) is solved after
entering each matrix element and it contains
 $2 \times \binom{7}{3} = 70$ inequality constraints                           
and 1-16 (= number of known entries in $\mathbf{D}$ ) equality constraints.
\[                                                                                      
\mathbf{D} =                                                                            
\begin{pmatrix}                                                                         
  1   &   3    &    9   &  3/2       &    6       &    5   &    2    \\                 
 1/3  &   1    &    3   &  1/2       &    2       &   3/2  &   1/2   \\                 
 1/9  &  1/3   &    1   &  \underline{1/6}       &   \underline{2/3}      &   1/2  &   1/5   \\                 
 2/3  &   2    &    \underline{6 }  &   1        &\underline{\mathbf{1/4}}&  \ast  &  \ast   \\                 
 1/6  &  1/2   &   \underline{3/2}  &\underline{\mathbf{4}}  &    1       &  \ast  &  \ast   \\                 
 1/5  &  2/3   &    2   & \ast       &  \ast      &   1    &  \ast   \\                 
 1/2  &   2    &    5   & \ast       &  \ast      &  \ast  &    1    \\                 
\end{pmatrix}                                                                           
\]                                                                                      

\begin{center}
\begin{tabular}{|c|c|c|c|c|c|c|c|c|c|c|c|}
\hline
             & $d_{1i}$                 &          &          &          &          &          &          &          &          &          &          \\
             &\tiny{$(i=2,3,\ldots,7)$} & $d_{23}$ & $d_{24}$ & $d_{25}$ & $d_{26}$ & $d_{27}$ & $d_{34}$ & $d_{35}$ & $d_{36}$ & $d_{37}$ & $d_{45}$ \\
\hline
$CM^{\ast}$  &   0                      &     0    &     0    &    0     &    1/10  &   1/4    &   1/4    &    1/4   &    1/4   &   1/4    &   $\mathbf{15/16 }$ (!)  \\
\hline
\end{tabular}
\end{center}

\medskip

Although the $CM$ threshold of acceptability for $7 \times 7$ matrices
has not been defined yet, 1/3 is assumed to be applicable again.
As $CM^{\ast} $ exceeds 1/3 (significantly)
when $d_{45} = \mathbf{1/4}$ is entered,
the inconsistency control identifies it as a possible mistype
and asks the decision maker for verification.
If the decision maker finds $d_{45} = \mathbf{1/4}$ appropriate
and the most inconsistent triad in unique, then
an error might occur before typing $d_{45}.$
In our example the most inconsistent triad in not unique, there are
three of them, one is underlined above. However, all three triads
contains $d_{45}.$ This fact suggests that the root of high
inconsistency is in $d_{45} = \mathbf{1/4}.$   \\                                                                        

The LP optimization problem (\ref{eq:a11}) has been implemented
in Maple 13 by using command \emph{LPSolve}
on a personal computer with 3.4 GHz processor and 2 GB memory.
CPU time, measured by command \emph{time() }, remains under 0.05 seconds
as the size of matrices varies between $3 \times 3$ and $ 9 \times 9$
and the number of missing elements varies between 1 and 28.

\section{Conclusions and final remarks}

In this study, we have demonstrated that the distance-based
inconsistency can be used to handle incomplete PC
matrices as a natural extension of the complete case. It is shown
that the determination of the minimal inconsistency level of the possible
extensions of an incomplete PC matrix is equivalent
to a linear programming problem.

In real life, gathering complete data may be difficult or time-consuming.
Incomplete PC matrices occur in each step of the filling in process even
the PC matrix is completely given in the end.
Thus the proposed approach is a useful tool for
signalling the impossibility of completion the given incomplete
matrix for an assumed inconsistency threshold as well as
for improving the level of inconsistency.

\section{Acknowledgments}
The authors thank the anonymous referees for their        
constructive reviews, especially the suggestions          
regarding to the scope of the paper.                      
This research has been supported
in part by NSERC grant in Canada and by OTKA grants K 60480, K 77420 in Hungary.

%
%

\end {document}